\documentclass{aa}  

\usepackage{graphicx}
\usepackage{txfonts}
\usepackage{comment}
\usepackage{bm}
\usepackage{braket}
\usepackage{xcolor}
\usepackage{booktabs}
\usepackage{adjustbox}
\usepackage[pdfencoding=auto,psdextra]{hyperref}
\hypersetup{
    colorlinks=true,
    linkcolor=blue,
    filecolor=magenta,      
    urlcolor=blue,
    citecolor=blue
}
\usepackage{multirow}
\usepackage[switch, modulo]{lineno}
\makeatletter
\renewcommand*\aa@pageof{, page \thepage{} of \pageref*{LastPage}}
\makeatother

\titlerunning{Denoising cosmological covariance matrices with Rotational Invariant estimators}
\authorrunning{A. Farina et al.}

\newcommand{\maT}[1]{\hat{\bm{#1}}}
\newcommand{\eigvec}[1]{\bm{\mathrm{#1}}}

\begin{document}

   \title{Denoising clustering covariance matrices with Rotational Invariant Estimators}


   \author{A. Farina\inst{1, 2, 3}
          \and
          M. Guidi\inst{4}
          \and
          A. Veropalumbo\inst{1, 2, 3}
          \and
          C. Guida\inst{2, 3}
          }

        \institute{
        INAF - Osservatorio Astronomico di Brera, Via Brera 28, 20122 Milano, Italy; via E. Bianchi 46, 23807 Merate, Italy\\
        \email{antonio.farina@inaf.it}
        \and
        Dipartimento di Fisica, Università degli Studi di Genova, Via Dodecaneso 33, I-16146 Genova, Italy
        \and
        INFN - Sezione di Genova, Via Dodecaneso 33, I-16146 Genova, Italy
        \and
        Dipartimento di Fisica e Astronomia Augusto Righi, Università di Bologna, Via Piero Gobetti 93/2, I-40129 Bologna, Italy
        }

   \date{}

 
\abstract
  {Cosmological parameter inference from galaxy clustering relies critically on accurate estimates of the covariance and precision matrices. In practice, these are often obtained from a limited number of mock catalogs, introducing noise and bias in the precision matrix when the data-vector dimension becomes comparable to the number of available realizations.}
  {We present the first application of the Rotational Invariant Estimator (RIE) to the large-scale clustering of galaxies, benchmarking it against the standard sample covariance and the non-linear shrinkage estimator NERCOME for both the two-point correlation function (2PCF) and power spectrum.}
  {Using controlled synthetic data sets with analytically known covariance matrices, we estimate the covariance with all three methods across a range of mock-to-dimension ratios $q = N/D$ and data-vector sizes $D$. We then perform Bayesian inference with an EFT-based model and quantify each estimator through the Figure of Bias (FoB) and Figure of Merit (FoM).}
  {After correction for finite-$N$ effects, the sample covariance recovers unbiased average uncertainty volumes but suffers from growing best-fit scatter and bias at small $q$ due to the Dodelson--Schneider effect. Both NERCOME and RIE substantially reduce these stochastic shifts; however, the amplitude of the uncertainties that they assign to the inferred parameters is probe-dependent. In configuration space, both estimators can yield overly tight constraints, with a bias that grows with $D$. In Fourier space, RIE delivers markedly improved best-fit stability with only a mild FoM bias, whereas NERCOME tends to overestimate the constraining power.}
  {Among the estimators tested, RIE emerges as the most effective at stabilizing best-fit recovery, particularly in Fourier space, where it closely reproduces the reference posteriors even when the number of mocks barely exceeds the data-vector dimension. However, its impact on posterior volumes must be carefully validated for each statistic and data-vector definition.}


   \maketitle
%

\section{Introduction}
\label{sec:intro}
In the era of precision cosmology, ensuring the reliability of statistical analysis is crucial to obtain unbiased estimates of cosmological parameters.
\noindent The standard approach to infer these parameters involves comparing theoretical models with data by sampling the posterior probability, generally under the assumption that the likelihood function follows a multivariate Gaussian distribution. In this procedure, the covariance matrix $\bm{C}$ and, in particular, its inverse, the precision matrix $\bm{C}^{-1}$, play a pivotal role. Indeed, they encapsulate the statistical uncertainty and correlations inherent in the data, thereby directly influencing the accuracy and reliability of parameter estimation.
Typically, the true data covariance is unknown, and so analyses are conducted utilising its sample estimate derived from simulated datasets

\begin{equation}
    \hat{M}_{ij} = \frac{1}{N}\sum_{k=0}^{N}{\left(d_i^k-\bar{d}_i\right)\left(d_j^k-\bar{d}_j\right)}\,,
    \label{eq:covariance}
\end{equation}

\noindent where $N$ is the number of mock
realizations, $d_i^k$ represents the element of
the data vector corresponding to the $i$-th bin
of the $k$-th catalogue, and $\bar{d}_i$ the average among the mocks.
The sample covariance estimator is unbiased, meaning that its expectation value always corresponds to the true covariance, regardless of the relative sizes of $N$ and $D$, where $D$ represents the length of the data vector. However, this property does not hold for the precision matrix.
Indeed, well-known statistical results (see e.g. \citealt{Anderson:2003}), which have been further discussed and developed in cosmology by several works (\citealt{Hartlap:2006, Taylor:2013, Dodelson:2013, Taylor:2014, Percival:2014, Sellentin:2015, Percival:2022}, among the others), show that
as $D$ approaches $N$ the precision matrix
becomes more and more biased, up to the limiting
case where $N=D +2$ and the covariance cannot be
inverted.\\
To tackle this 'dimensionality curse'
different approaches have been adopted in galaxy clustering literature. First, purely theoretical covariance matrix models can be utilized, which in the case of $N$-point correlation functions (NPCF) can be obtained within the framework of Perturbation Theory (PT) techniques \citep{Grieb:2016, Lippich:2019, Li:2019, Sugiyama:2020, Wadekar:2020, WadekarS:2020}.
However, these models' applicability is basically restricted to the linear regime of cosmic structure formation. In the non-linear regime, where gravitational interactions become complex, PT often falls short.
To overcome, at least partially, this issue, purely theoretical predictions can be conveniently integrated with additional free parameters that can be calibrated using a limited number of simulations. This hybrid approach offers greater flexibility and has been shown to be effective on a wide range of scales and statistics (see, e.g., \citealt{Pearson:2016, O'Connell:2016, O'Connell:2019, Philcox:2019, Philcox:2020, Colavincenzo:2020, Fumagalli:2022}).
Alternative approaches aim to mitigate the breakdown of the regime $N \gg D$ by exploiting data compression schemes (e.g. \citealt{Philcox:2021b}), reducing the impact of covariance noise via resampling techniques such as jackknife and bootstrap methods \citep{Norberg:2009, Escoffier:2016, Howlett:2017, Klypin:2018, Mohammad:2022}, or, more recently, by using machine learning algorithms to identify and suppress noise in sample covariance estimates \citep{DeSanti:2022}. \\
In this work, instead, we follow \cite{Pope:2008, Paz:2015, Padmanabhan:2016, Friedrich:2016} and \cite{Joachimi:2017} and consider alternative ways to estimate cosmological precision matrices from simulations.\\
In particular, in this work we employ for the first time on a cosmological dataset, the Rotational Invariant Estimator (RIE) proposed by \cite{Bun:2016}. Unlike NERCOME, which operates by mixing covariance estimates from independent subsets, RIE leverages an analytical denoising of the eigenvalue spectrum rooted in Random Matrix Theory, making it potentially better suited to high-dimensional problems where the spectral structure of the covariance plays a dominant role.\\
We will assess RIE's performance in the realm of galaxy clustering parameter estimation by comparing it with the non-linear shrinkage estimator NERCOME \citep{Lam:2016,Joachimi:2017}.
To achieve this, we analyze several synthetic datasets of varying dimensionality, corresponding to both galaxy two-point correlation function (2PCF) and power spectrum, for which the true covariance matrices are precisely known. This allows for a controlled and quantitative comparison of different covariance estimators.\\
This paper is organized as follows. In Sec.~\ref{sec:estimators}, we describe the sample, RIE, and NERCOME covariance estimators and detail their implementation. Sec.~\ref{sec:models} presents the theoretical models used to describe the 2PCF and power spectrum of the synthetic datasets, which are introduced in Sec.~\ref{sec:datasets}. The statistical methods adopted in our analysis are summarized in Sec.~\ref{sec:likelihood}. The main results are discussed in Sec.~\ref{sec:results}, and we conclude in Sec.~\ref{sec:conclusion} with a summary and perspectives for future work.

\section{Covariance matrix estimators}\label{sec:estimators}

\subsection{Sample estimator}
Despite being unbiased, the sample estimator presented in Eq.~\eqref{eq:covariance} can be problematic in parameter inference when only a limited number of realizations is available. In particular, when estimating the precision matrix $\mathbf{C}^{-1}$ by directly inverting the sample covariance, two distinct effects arise.\\
The first is a systematic bias. For Gaussian-distributed data, the sample covariance follows a Wishart distribution \citep{Wishart:1928}, and its inverse therefore follows an inverse Wishart distribution (see e.g. \citealt{Anderson:2003}). The expectation value of the naive precision estimator $\maT{M}^{-1}$ is then given by (see e.g. \citealp{Kaufman:1967, VonRosen:1988}):

\begin{equation}
\langle \maT{M}^{-1} \rangle = \frac{N - 1}{N - D - 2}\, \mathbf{C}^{-1}\,.
\end{equation}

\noindent This expression shows that $\maT{M}^{-1}$ systematically overestimates the true precision matrix when $N$ is not much larger than $D$. The bias can be corrected by introducing an unbiased estimator,

\begin{equation}
\maT{M}_U^{-1} = \frac{N - D - 2}{N - 1}\, \maT{M}^{-1}\,,
\end{equation}

\noindent a correction originally proposed in a cosmological context by \cite{Hartlap:2006}, and now widely known in the field as the Hartlap factor.\\
The second effect arises from sampling noise. Even after correcting for the systematic bias, a finite number of realizations introduces random fluctuations, $\Delta\mathbf{C}^{-1}$, around the true precision matrix, such that $\maT{M}^{-1} = \mathbf{C}^{-1} + \Delta\mathbf{C}^{-1}$. For Gaussian-distributed data, the covariance of the estimated precision matrix can be expressed analytically \citep{Taylor:2013} as

\begin{equation}
\langle \Delta C_{ij}^{-1} \Delta C_{k\ell}^{-1} \rangle = A \,C_{ij}^{-1} C_{k\ell}^{-1} + B \, (C_{ik}^{-1} C_{j\ell}^{-1} + C_{i\ell}^{-1} C_{jk}^{-1})\,,
\end{equation}

\noindent where 

\begin{equation}
\begin{split}
A = &\frac{2}{(N - D - 1)(N - D - 4)}\\
B = &\frac{N - D - 2}{(N - D - 1)(N - D - 4)}.
\end{split}
\end{equation}

\noindent These stochastic fluctuations propagate into parameter inference in three main ways: they increase the variance of the estimated parameter uncertainties \citep{Taylor:2013}, introduce random shifts in the location of the best-fit parameters \citep{Dodelson:2013}, and bias the average size of the inferred parameter errors \citep{Percival:2014}.
To account for these effects, \cite{Percival:2014} proposed a multiplicative correction factor

\begin{equation}
m_1 = \frac{1 + B\,(D - N_p)}{1 + A + B\,(N_p + 1)}\,,
\end{equation}

\noindent where $N_p$ is the number of free parameters in the analysis. When applied to the parameter variances, this factor compensates for the statistical noise introduced by covariance estimation, ensuring that the reported uncertainties more accurately reflect the true parameter dispersion. Related approaches have been developed by \cite{Sellentin:2015} and \cite{Percival:2022}, in which the effect of a finite number of realizations is incorporated by modifying the likelihood function from a Gaussian to a multivariate $t$-distribution. This effectively accounts for both bias and sampling noise in a single framework. In the present work, however, we restrict our analysis to Gaussian likelihoods, and therefore rely on the standard Hartlap and Percival corrections to account for bias and finite-sample noise in the precision matrix.

\subsection{Rotational Invariant Estimator}
Let us consider the statistical problem of determining the true data covariance $\bm{C}$ given its noisy sample realization $\hat{\bm{M}}$. In particular, in this section we focus on the class of Rotational Invariant Estimators (RIE), 
defined as the ensemble $\maT{\Xi}(\maT{M})$ of matrix-valued functions of $\maT{M}$ satisfying

\begin{equation}
    \bm{U}\,\maT{\Theta}\,\bm{U}^T = \maT{\Theta}(\bm{U}\maT{M}\bm{U}^T)
\end{equation}

\noindent for every $\maT{\Theta}\in\maT{\Xi}(\maT{M})$ and every $\bm{U}\in \mathcal{O}(N)$, 
where $\mathcal{O}(N)$ denotes the group of $N\times N$ orthogonal matrices.
Under this condition, every estimator $\maT{\Theta}\in\maT{\Xi}(\maT{M})$ shares the same 
eigenvector basis as $\maT{M}$. In what follows we will always refer to the eigenvectors of a 
generic $\maT{\Theta}\in\maT{\Xi}(\maT{M})$ as $\eigvec{u}_i\;\forall\, i = 1,.., D$, while 
with $\eigvec{v}_i$ we will indicate those of the true covariance matrix. The optimal estimator 
$\maT{\Pi}\in\maT{\Xi}(\maT{M})$ of $\bm{C}$ is found by solving the minimization problem:

\begin{equation}
    \maT{\Pi} = \underset{\maT{\Theta}\,\in\,\maT{\Xi}(\maT{M})}{\mathrm{argmin}}\, 
    \lVert\maT{\Theta} - \bm{C} \rVert
\end{equation}

\noindent where $\lVert\maT{\Theta} - \bm{C}\rVert \equiv 
Tr\left[\left(\maT{\Theta} - \bm{C}\right)^2\right]$ is the standard $\mathbb{L}_2$ norm. 
Projecting on the eigenbasis, one finds that the eigenvalues $\lambda_i$ of $\maT{\Pi}$ 
can be expressed as:

\begin{equation}
    \lambda_i = \sum_{j}{c_j\left(\eigvec{u}_i\cdot\eigvec{v}_j\right)^2}
\end{equation}

\noindent where $c_j$ are the eigenvalues of the true covariance matrix.\\ The interpretation of this result is clear: the choice of a wrong, noisy eigenbasis is explicitly corrected by the RIE by weighting the true eigenvalues $c_j$ with the squared scalar products between the noisy eigenvectors and the true ones. This products, often referred to as overlaps, encode information about the relative directions spanned by the two eigenbasis. 
Involving eigenvalues and eigenvectors of the true covariance, the matrix $\maT{\Pi}$, often called the 'oracle' estimator, may seem impractical.
However, leveraging techniques coming from Random Matrix Theory (RMT), it can be shown that in the limit of large datasets, that is for $D\rightarrow\infty$, the 'oracle' eigenvalues self-average to a fully computable, analytical expression.
The full derivation of such expression is out of the scope of this manuscript so, for further details, we refer the reader to \cite{Bun:2016, Bun:2017, Bun:2018} and references therein, works which delve deep on the technical aspects of this kind of calculations. 
Instead, we directly jump to the final result which reads:

\begin{equation}
    \lambda_i(\xi_i) = \frac{\xi_i}{\left|1 + q - q\,\xi_i\, \lim_{\eta\rightarrow 0^+}\mathfrak{g}_{\maT{M}}(\xi_i - i\eta)\right|}
    \label{eq:RIE}
\end{equation}

\noindent where $\xi_i$ are the eigenvalues of the sample covariance $\maT{M}$, $q = D/N$ and $\mathfrak{g}_{\maT{M}}(z)$ is the Stieltjes transform of the resolvent $G_{\maT{M}}(z) = (z\bm{I} - \maT{M})^{-1}$, that is:

\begin{equation}
    \mathfrak{g}_{\maT{M}}(z) = \frac{1}{D}Tr\left[G_{\maT{M}}(z)\right]
\end{equation}

\noindent As anticipated, the right-hand side of Eq.~\eqref{eq:RIE} no longer involves the matrix $\bm{C}$ and depends solely on quantities that can be directly estimated from data.
We observe that, while in Eq.~\eqref{eq:RIE} there is a limit for $\eta$ that goes to zero, this quantity is also required to be bigger than $D^{-1}$ \citep{Bun:2016}. This double condition can be fully satisfied only for $D\rightarrow\infty$, a requirement never realised in practice. 
A good and largely adopted compromise consists of fixing $\eta = 1/\sqrt{D}$. We will align to this choice throughout this paper. 
Finally, in this work we avoid applying the RIE directly to the sample covariance matrix. Instead, we first apply it to the correlation matrix and only at the final stage we rescale the result to the original variances. This approach allows us to isolate the effects of regularization on the correlation structure while ensuring that the variances are correctly normalized in the final covariance matrix, preserving the statistical meaning of the data.

\subsection{NERCOME}
Finally, we consider the NERCOME (Non-parametric Eigenvalue Regularised Covariance Matrix Estimator) algorithm. First proposed by \cite{Lam:2016} and based on previous works of \cite{Ledoit:2012} and \cite{Abadir:2014}, NERCOME has been utilized on cosmological simulated datasets by \cite{Joachimi:2017, Beauchamps:2023} and by \cite{Looijmans:2024} on the BOSS DR12 data \citep{Alam:2015}.\\
The core idea of this approach is basically that mixing two covariance matrices estimated from independent datasets would help shrink the eigenvalue spectrum and, thus, lead to a well-conditioned covariance matrix.
Specifically, the NERCOME algorithm consists of three main steps:

\begin{enumerate}
    \item \emph{Subset Division}: split the set of realizations into two subsets. The first subset contains $s$ realizations, while the second contains $N-s$ realizations.    
    \item \emph{Covariance Estimation}: For each subset, apply the sample estimator to derive the covariance matrices $\maT{S}_i$ where $i=1,2$ depending on the particular subset. Each covariance matrix is then decomposed into its eigenvectors and eigenvalues in the form $\maT{S}_i=\bm{U}_i\,\bm{D}_i\,\bm{U}_i^T$ where $\bm{U}$ is the matrix of eigenvectors, and $\bm{D}$ is the diagonal matrix containing the eigenvalues.
    \item \emph{Covariance Matrix Estimation and Averaging}: Estimate the covariance matrix using the formula:

    \begin{equation}
        \maT{Z} = \maT{U}_1\,\mathrm{diag}\left(\maT{U}_1^T\,S_2\,\maT{U}_1\right)\,\maT{U}_1^T
    \end{equation}

    \noindent Finally, $\maT{Z}$ is averaged over $N_{R}$ different, random compositions of the two subsets while mantaining a fixed split position $s$.
\end{enumerate}

\noindent The optimal split position $s$ can then be found by minimising the distance $\lVert\bar{\maT{Z}}(s) -\bar{\maT{S_2}}(s) \rVert$, where the bar denotes the average over the $N_{R}$ compositions mentioned in point 2 \citep{Lam:2016}. In what follows, we always use results from the NERCOME implementation provided by \cite{Looijmans:2024}\footnote{\url{https://github.com/marnixlooijmans/shrinkage-estimation-paper}}, which automatically identifies and employs the optimal split position.


\section{Theoretical models}
\label{sec:models}
In this section, we introduce the theoretical models for the galaxy 2PCF and power spectrum, which serve to evaluate the performance of the covariance matrix estimators discussed in Sec.~\ref{sec:estimators}. 
To effectively model the power spectrum, we adopt the one-loop perturbative framework provided by the Effective Field Theory of Large-Scale Structure \citep[EFT of LSS,][]{Carrasco:2012, Senatore:2014, Assassi:2014, Lewandowski:2018}. Specifically, we write our power-spectrum model as:

\begin{equation}
\begin{split}
    P_{\mathrm{nlo}}(k,\mu) =\; &Z_1^2(\bm{k}, \mu)\;P_\mathrm{m}(k)\; +\\
             +&2\int{\frac{d^3q}{(2\pi)^3}\,Z_2^2(\bm{q}, \bm{k-q}, \mu)P_{\mathrm{m}}(|\bm{k-q}|)P_{\mathrm{m}}(q)} \;+\\
             +\;&6\,Z_1(\bm{k}, \mu)P_{\mathrm{m}}(k)\int{\frac{d^3q}{(2\pi)^3}Z_3(\bm{q}, \bm{-q}, \bm{k}, \mu)P_{\mathrm{m}}(q)}\; +\\
             -\;&2\left[c_0\,+\,c_2\,\mathcal{L}_2(\mu)\,\right]\,k^2P_{\mathrm{m}}(k)\;+\\
             -\;&c_{\mathrm{nlo}}f^4\mu^4Z_1(\bm{k}, \mu)\,k^4P_{\mathrm{m}}(k)
\end{split}    
    \label{eq:Pk_NoIR}
\end{equation}

\noindent The expression above involves several components. First of all, $\mathcal{L}_\ell(\mu)$ indicates the Legendre polynomial of order $\ell$, where $\mu=\bm{k}\cdot\hat n /k$ defines the cosine of the angle between the wavevector $k$ and the line-of-sight (LOS) $\hat n$. $P_{\mathrm{m}}(k)$, instead, indicates the linear matter power spectrum.
Second, the $Z_n$ terms are derived from Eulerian perturbation theory and incorporate effects such as non-linear structure growth, redshift-space distortions (RSD), and the galaxy bias relation. Detailed formulas for these kernels can be found in Appendix A.2 of \cite{Philcox:2022B}. 
Finally, the coefficients $c_0$, $c_2$, and $c_{\mathrm{nlo}}$ serve as EFT counter-terms. These terms address UV divergences in loop integrals and account for corrections arising from deviations in fluid dynamics (see e.g. \citealt{Carrasco:2012}). 
To account for the degradation of the Baryonic Acoustic Oscillation (BAO) signal driven by the non-linear evolution of cosmic structures \citep{Crocce:2008, Matsubara:2008, Matsubara:2008a}, in Eq.~(\ref{eq:Pk_NoIR}) we include the Infrared Resummation (IR) scheme proposed by \cite{Ivanov:2018}. For the details, we refer the reader to Eqs.(2.8)-(2.11) of \cite{Farina:2026}. Lastly, we include the Alcock-Paczynski (AP) effect \citep{Alcock-Paczynski:1979}, i.e. the impact of geometric distortions that arises from adopting an incorrect fiducial cosmology, by introducing the following mappings for $k$ and $\mu$:

\begin{equation}
    \begin{aligned}
    k \longrightarrow \;&k^\prime \equiv k\left[\left(\frac{H}{H_{\mathrm{fid}}}\right)^2\mu^2 + \left(\frac{D_{A, \,\mathrm{fid}}}{D_{A}}\right)^2 \left(1 - \mu^2\right)  \right]^{\frac12}, \\
    \mu \longrightarrow \;&\mu^\prime \equiv \mu \left(\frac{H}{H_{\mathrm{fid}}}\right)\, \left[\left(\frac{H}{H_{\mathrm{fid}}}\right)^2\mu^2 + \left(\frac{D_{A, \,\mathrm{fid}}}{D_{A}}\right)^2 \left(1 - \mu^2\right)  \right]^{-\frac12}
    \end{aligned}
    \label{eq:AP_maps}
\end{equation}

\noindent where the subscript "$\mathrm{fid}$" denotes the fiducial Hubble parameter, $H$, and angular diameter distance $D_A$.
We decompose the power-spectrum in Legendre polynomials and consider in our analysis only its first three multipole moments, $\ell =0,\,2$ and 4.\\
The 2PCF multipoles are obtained as the Hankel transform of their power-spectrum counterparts, computed using the FFTLog algorithm \citep{Hamilton:2000}. In practice, both the power spectrum and the 2PCF are computed using the \texttt{MElCorr} code \citep{Farina:2026}. To accelerate the parameter inference, we first use \texttt{MElCorr} to generate a training set of power spectra and correlation functions, which is then employed to train an emulator. The emulator is subsequently used in place of direct \texttt{MElCorr} evaluations, allowing for a much faster parameter space sampling. Details of the emulator construction and training procedure are provided in Appendix \ref{app:emulator}.

\section{Datasets}
\label{sec:datasets}
To evaluate the performance of our covariance matrix estimators in a controlled setting, we construct an ideal scenario for both the 2PCF and power spectrum, where the true covariance of the dataset is known a priori.
Starting from a model realization $m$ of the probe under consideration and its Gaussian covariance matrix $\bm C$, we generate the $k$-th dataset realization as

\begin{equation}
    d^k = m + \sqrt{\bm C}X^k\,,
    \label{eq:cholesky}
\end{equation}

\noindent where $\sqrt{\bm C}$ is the Cholesky decomposition of $\bm C$ and $X^k$ is a normal random vector. In particular, we compute the Gaussian covariance matrix $\bm C$ as described in \cite{Grieb:2016} and implemented in the \texttt{GaussianCovariance} package\footnote{\href{https://gitlab.com/veropalumbo.alfonso/gaussiancovariance}{https://gitlab.com/veropalumbo.alfonso/gaussiancovariance}}, assuming a survey volume of $V = 3.175\ h^{-3}\mathrm{Gpc}^3$ and a mean galaxy number density of $\bar{n} = 2.13\, h^3\mathrm{Mpc}^{-3}$. For each dataset, we generate 500,000 noisy realizations of the probe for covariance matrix estimation, along with an additional realization to serve as data vector.
To generate such datasets, we assume $z=1.317$ and consider a \textit{DESI}+\textit{Planck}+\textit{Union3}-like cosmology \citep{Adame:2025} whose parameters are listed in Tab.~\ref{tab:cosmo_pars}.

\begin{table}[t]
\centering
\caption{Fiducial DESI-like cosmology adopted in this work.}
\label{tab:cosmo_pars}
\begin{tabular}{lc}
\hline\hline
Parameter & Value \\
\hline
$z_{\mathrm{eff}}$ & $1.317$ \\
$h$ & $0.6736$ \\
$\Omega_b$ & $0.0493$ \\
$\Omega_m$ & $0.3137$ \\
$n_s$ & $0.9649$ \\
$A_s$ & $2.19\times10^{-9}$ \\
$m_\nu$ [eV] & $0.06$ \\
$(w_0,\,w_a)$ & $(-0.727,\,-1.05)$ \\
\hline
\end{tabular}
\end{table}

\noindent The nuisance parameters, instead, are chosen arbitrarily and summarized in Tab.~\ref{tab:nuisance_pars}.

\begin{table}[t]
\centering
\caption{Fiducial nuisance parameters adopted in this work.}
\label{tab:nuisance_pars}
\begin{tabular}{lc}
\hline\hline
Parameter & Value \\
\hline
$b_1$ & 1.92 \\
$b_2$ & 0.4 \\
$b_{\mathcal{G}_2}$ & -0.4 \\
$b_{\Gamma_3}$ & 0.1 \\
$c_0$ & -4 \\
$c_2$ & 48 \\
$c_4$ & 10 \\
$c_{nlo}$ & 0.0 \\
\hline
\end{tabular}
\end{table}

\noindent The power spectrum is computed in the interval from $k_{\rm min} = 0.01~h\,{\rm Mpc}^{-1}$ to $k_{\rm max} = 0.3~h\,{\rm Mpc}^{-1}$, while the 2PCF is computed over $r_{\rm min} = 20~{\rm Mpc}/h$ to $r_{\rm max} = 200~{\rm Mpc}/h$. To explore the effect of covariance matrix size on the RIE method, we consider three different data-vector lengths, namely $D = 90,~ 180,$ and 270 bins.

\section{Parameter inference}
\label{sec:likelihood}
Given models and datasets, we infer the values and uncertainties of cosmological parameters $\bm \theta$ by sampling the posterior probability $P(m(\bm \theta)\,|\,\bm d)\propto \mathcal{L}(\bm d\,|\,m(\bm \theta))\,P(\bm\theta)$, where $m(\bm\theta)$ is the model prediction for a certain set of parameters, $P(\bm\theta)$ the prior probability and $\mathcal{L}(\bm d\,|\,m(\bm \theta))$ the likelihood function, that we assume to be Gaussian. Under this assumption:

\begin{equation}
    \log{\mathcal{L}} = -\frac12\left(\bm d-\bm \mu(\bm\theta)\right)^T\maT{C}^{-1}\left(\bm d-\bm \mu(\bm\theta)\right).
\end{equation}

\noindent In our analysis, we vary 2 cosmological and 4 galaxy bias parameters other than 3 EFT counterterms:

\begin{equation}
    \left(w_{cdm},\;h\right)\times\left(b_1, \;b_2,\;b_{\mathcal{G}_2}, b_{\Gamma_3}\;\right) \times\left(c_0,\;c_2,\;c_{\mathrm{nlo}}\right).
\end{equation}

\noindent We fix all the remaining cosmological parameters to their fiducial values (Tab. \ref{tab:cosmo_pars}).
For all free parameters, we adopt flat priors; their bounds are listed in Tab.~\ref{tab:prior_intervals}.

\begin{table}[t]
\centering
\caption{Flat prior intervals adopted for the free parameters of the model.}
\label{tab:prior_intervals}
\begin{tabular}{lc}
\hline\hline
Parameter & Prior interval \\
\hline
 $h$  & $[0.55,\,0.85]$ \\
$w_c$ & $[0.11,\,0.41]$ \\
$b_1$ & $[1,\,5.0]$ \\
$b_2$ & $[-10,\,10]$ \\
$b_{\mathcal{G}_2}$ & $[-10,\,20]$ \\
$b_{\Gamma_3}$ & $[-30,\,20]$ \\
$c_0$ & $[-100,\,100]$ \\
$c_2$ & $[-100,\,100]$ \\
$c_{nlo}$ & $[-1000,\,1000]$ \\
\hline
\end{tabular}
\end{table}

\noindent We sample the posterior distribution using the importance nested sampling method implemented in \texttt{Nautilus} \citep{Lange:2023}, which uses neural networks to accelerate the sampling. We adopt $n_{\rm live}=2000$ live points and run the sampler until the estimated remaining contribution to the Bayesian evidence falls below the built-in tolerance, i.e., $\Delta \log Z < 0.01$.

\section{Results}
\label{sec:results}
In this section, we investigate the impact of sampling noise in the covariance matrices of the 2PCF and power spectrum on parameter inference, following the approach of \cite{Beauchamps:2023}. We fix a single realization of the data vector for each probe and repeat the parameter inference using different estimates of the covariance matrix obtained from independent sets of synthetic realizations. This setup isolates the effect of noise in the covariance estimate on the inferred parameters.
To study the dependence on the number of realizations used to estimate the covariance, we vary the number of mocks $N$, considering sets containing $q \times D$ realizations with $q \in [1.2,~2,~4,~8,~16]$, where $D$ is the length of the data vector. For each value of $N$, we generate 50 independent sets of realizations and estimate a covariance matrix from each set. Repeating the inference with these covariance matrices while keeping the data vector fixed yields a distribution of best-fit parameters and uncertainties that quantifies the systematic bias and statistical scatter induced by the finite number of mocks.
To evaluate whether the covariance estimators discussed in Sec.~\ref{sec:estimators} introduce biases in the best-fit estimates or their associated uncertainties, we adopt two complementary metrics: the Figure of Bias (FoB) and the Figure of Merit (FoM). The FoB measures the deviation of the best-fit parameters, $\hat{\theta}$, from their true values, $\theta_\mathrm{true}$, and is defined as

\begin{equation}
    \mathrm{FoB} = \left[\left(\hat{\bm\theta} - \bm{\theta}_\mathrm{true}\right)^T \maT{C}_p^{-1} \left(\hat{\bm\theta} - \bm{\theta}_\mathrm{true}\right)\right]^{1/2}\,,
    \label{eq:FoB}
\end{equation}

\noindent where $\maT{C}_p$ is the covariance matrix of the estimated parameters. The FoM, instead, quantifies the precision of the parameter constraints and is expressed as the inverse of the volume enclosed by the confidence region in parameter space:

\begin{equation}
    \mathrm{FoM} = \frac{1}{\sqrt{\mathrm{det}(\maT{C}_p)}}\,.
    \label{eq:FoM}
\end{equation}

\noindent In the following, we split the analysis into two distinct parts: in Sec.~\ref{sec:2PCF_analysis} we focus on results obtained from the 2PCF, while in Sec.~\ref{sec:Pk_analysis} we present the corresponding Fourier space analysis. Finally, in all the plots presented in this section, unless otherwise specified, results labeled as "sample covariance" correspond to analyses performed using the inverse sample covariance corrected by the Hartlap factor only. The additional inflation of the parameter uncertainties introduced by the Percival factor, which accounts for the increased scatter of best-fit parameters when the covariance is estimated from a finite number of realizations, is not included here for clarity.

\subsection{2PCF results}
\label{sec:2PCF_analysis}

\begin{figure}[t] 
    \centering
    \includegraphics[width=\columnwidth]{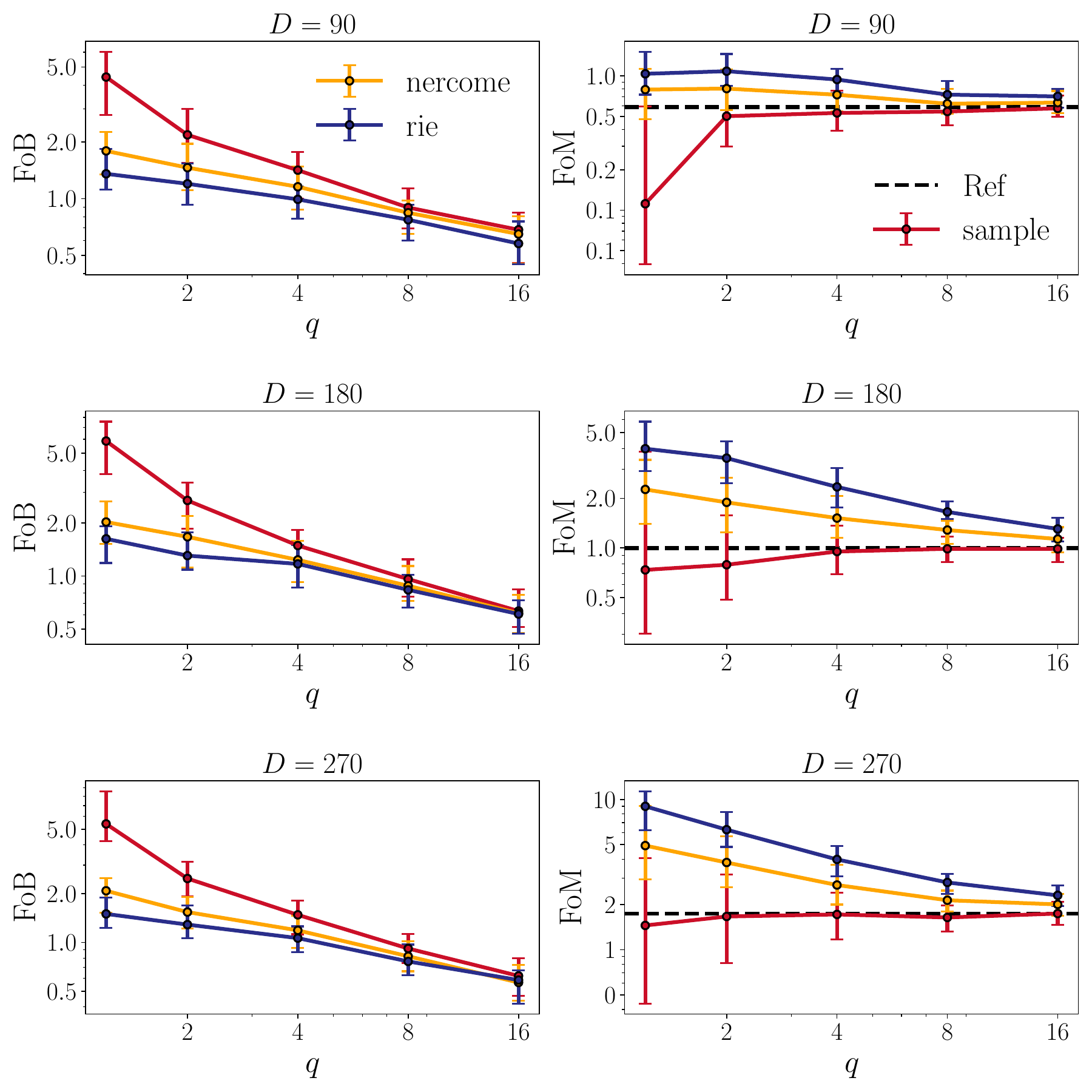}
    \caption{\small \emph{Left panels:} FoB of the 2PCF of the synthetic dataset as a function of $q$ for the sample (red), NERCOME (gold) and RIE (blue) estimators. Dots and error bars are, respectively, mean and \emph{rms} scatter of the FoB as obtained from 100 different realizations of the covariance matrix. Each row correspond to a different datavector length $D$ as reported in the panel titles. \emph{Right panels:} Same plots but for the FoM. The horizontal, black dashed line represent the FoM associated to the true covariance matrix.}
    \label{fig:fom_fob_vs_q_multi_bins}
\end{figure}

The main results of the 2PCF analysis are summarized in Fig.~\ref{fig:fom_fob_vs_q_multi_bins}. For each choice of the data vector size $D$, the figure displays the median FoB (left panels) and FoM (right panels) as a function of $q$, along with their 68\% confidence regions (error bars). Results are shown for the sample (red), NERCOME (gold), and RIE (blue) covariances.

\begin{figure}[htbp]
    \centering
    \includegraphics[width=\columnwidth]{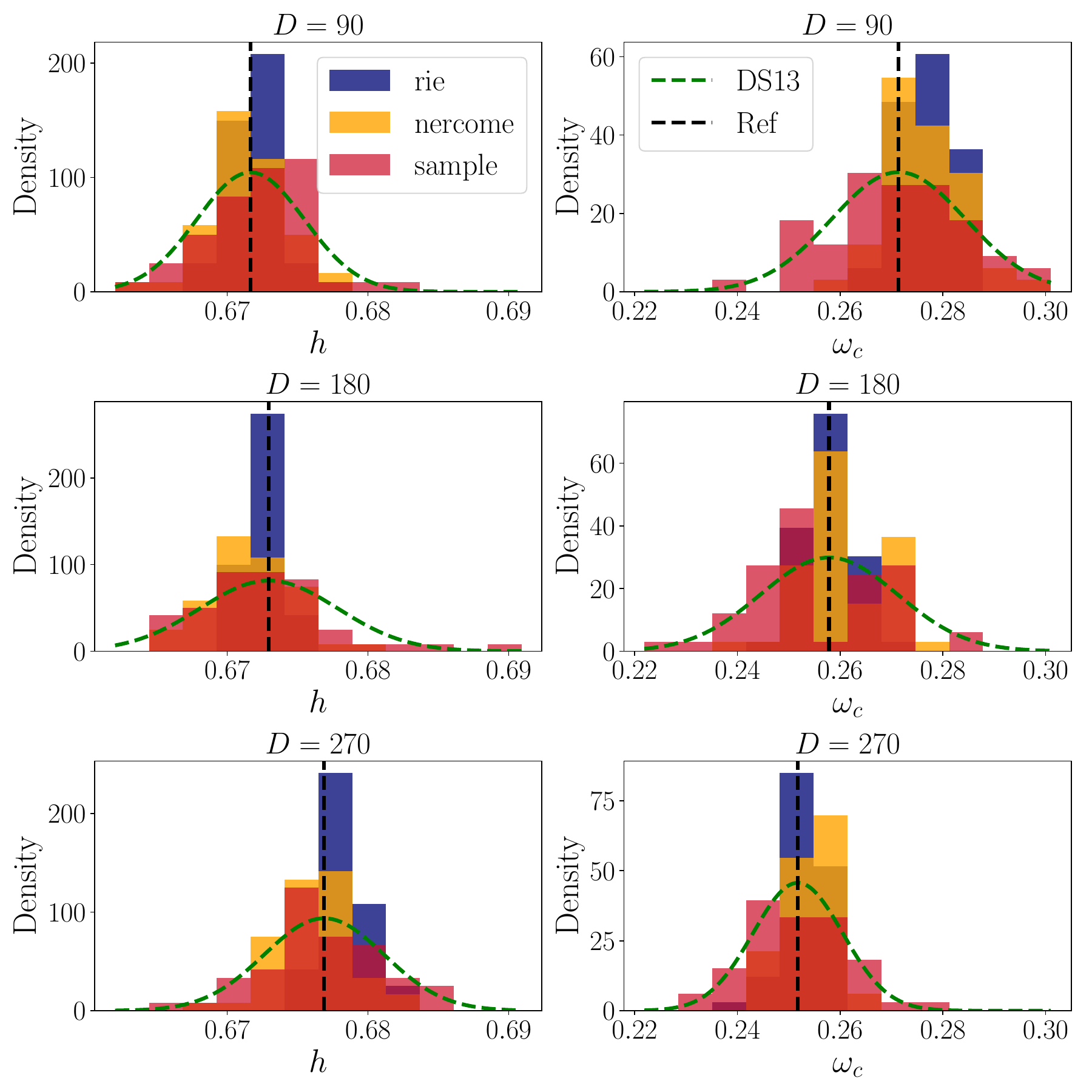}
    \caption{\small Normalised distributions of the best-fit parameter estimates for $h$ (left column) and $\omega_c$ (right column) derived from the 2PCF for different data vector sizes $D$.
    Results are shown for NERCOME (gold), RIE (blue), and sample covariance (red). 
    The vertical black dashed lines indicate the true parameter values inferred from the exact covariance, while the cyan dashed lines show the distribution of best-fit values predicted by DS13.}
    \label{fig:distros}
\end{figure}

\begin{figure*}[htbp]
    \centering
    \includegraphics[width=\textwidth]{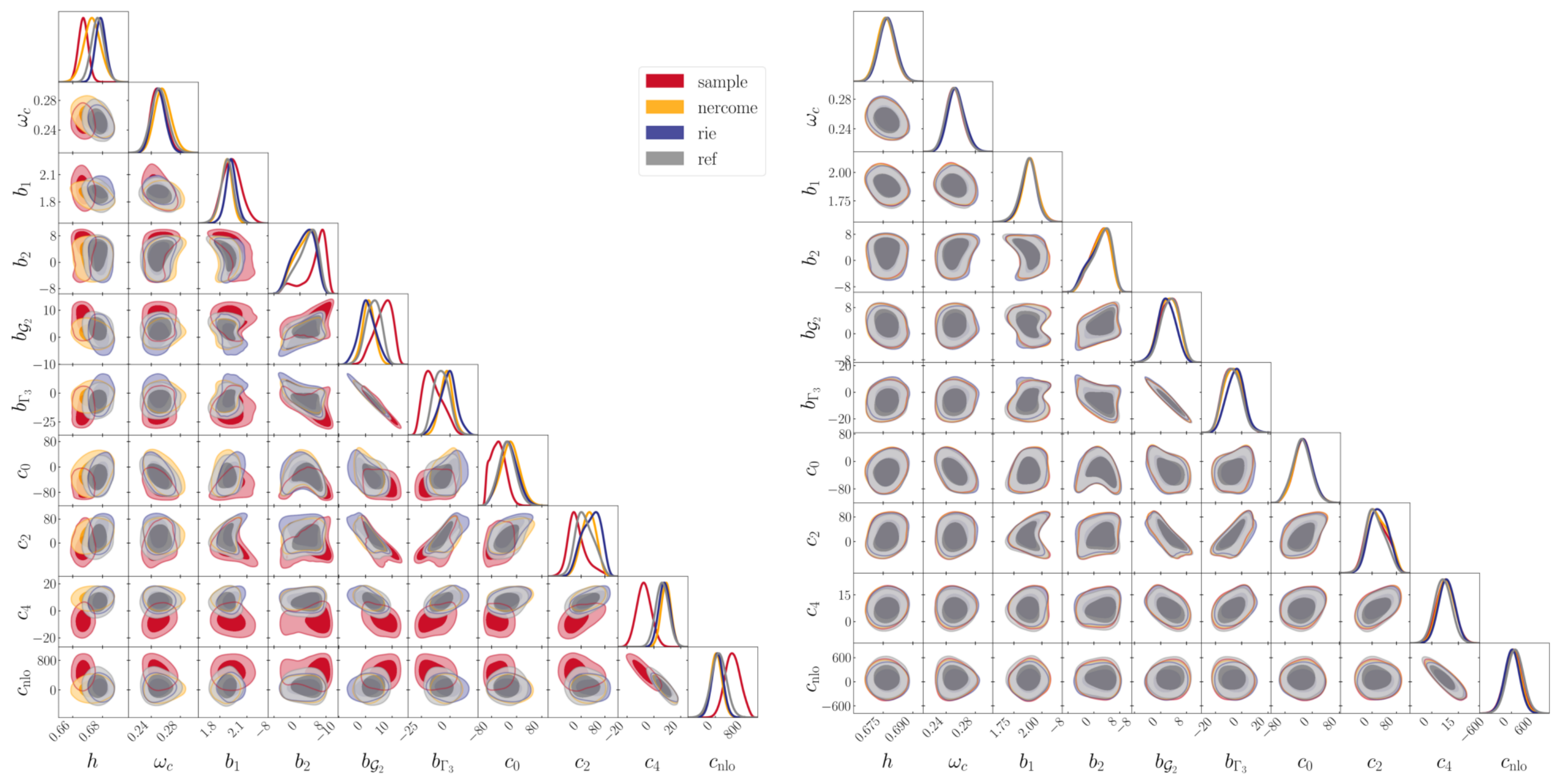}
    \caption{\small Posterior distributions for the model parameters obtained from the 2PCF analysis for $D=270$ using different covariance matrix estimators. For each method, we show the realization whose FoM and FoB are closest to the median values of their respective distributions. Results are shown for two values of $q$ ($q=1.2$ and $q=16$), comparing sample (red), NERCOME (gold), RIE (blue), and the reference case (grey).}
    \label{fig:triangle_median_real}
\end{figure*}

\noindent For all the explored $D$ values, the best-fit estimates obtained using the three different covariance estimators remain consistent down to 
$q=2$. Below this threshold, the sample covariance begins to deviate from the others, exhibiting increasingly biased predictions.
Notably, not only is the mean value of the FoB systematically higher for the sample estimator compared to the NERCOME and RIE cases, but the associated error bars are also significantly larger for small values of $q$.
This behavior is due to the well-known Dodelson-Schneider effect (\citealt{Dodelson:2013}, DS13 hereafter), for which noisy estimates of the covariance matrix lead to stochastic shifts of the best-fit estimates. This is further illustrated in Fig.~\ref{fig:distros}, which shows the normalized distributions of the best-fit values of $h$ and $\omega_c$ at fixed $q=2$ for the three $D$ values considered. Only these two parameters are shown for illustration purposes but analogous results are obtained for all the other free parameters of the model. As evident, the sample covariance distributions are in good agreement with the predictions of DS13, represented by the green dashed line, in most of the cases. On the contrary, NERCOME and RIE appear to be peaked around the correct parameter values estimated using the true covariance (vertical, black dashed line) but show distributions less broadened than the sample estimator,
indicating that these two methods are significantly less affected by the Dodelson-Schneider effect\footnote{Note that, regarding NERCOME, this result confirms the findings of \cite{Beauchamps:2023}, which however performed their analysis in Fourier space.}.
In the right panels of Fig.~\ref{fig:fom_fob_vs_q_multi_bins}, we present the median and 68\% uncertainty region of the FoM as a function of $q$ for the three covariance matrix estimators analyzed in this work. As for the the left column of the Figure, each row refers to a different $D$ value. The horizontal black dashed line indicates the correct FoM derived using the true covariance matrix.
Although the sample covariance (which we recall is corrected here solely with the Hartlap factor) yields a highly accurate estimate of the credible volume in parameter space, both NERCOME and RIE systematically underestimate the associated uncertainties, leading to an inflated FoM relative to the true value. NERCOME performs reasonably well at $D=90$, but its bias grows progressively as $D$ increases. RIE exhibits this trend even more strongly: it is already biased at $D=90$ and becomes increasingly inaccurate for larger $D$ values.\\
These results diverge significantly from those of earlier studies on NERCOME, which have consistently shown that this estimator tends to overestimate uncertainties in parameter inference. However, previous works highlighted this behavior in the context of weak lensing 2PCF \citep{Joachimi:2017} and the galaxy power spectrum \citep{Beauchamps:2023, Looijmans:2024}. Therefore, to investigate this discrepancy further, in the next section we extend the analysis presented in this section for the galaxy 2PCF to the corresponding power spectrum.\\
Before proceeding, we illustrate the posterior distributions visually. Since the behavior of the covariance matrix estimators is stochastic, for display purposes we select, for each of them, the realization whose FoM and FoB are closest to the median values reported in Fig.~\ref{fig:fom_fob_vs_q_multi_bins}. We show the resulting contour plots for $D=270$ and two values of $q$, namely 1.2 and 16, in Fig.~\ref{fig:triangle_median_real}.

\noindent For $q=1.2$ (left panel), the differences among estimators are striking: the sample contours (red) are not only significantly broader and shifted relative to the reference (grey), but in some parameter subspaces they are visibly bounded by the prior edges or exhibit different degeneracy directions, features that bear little resemblance to the well-behaved reference posteriors. The NERCOME (gold) and RIE (blue) contours sit considerably closer to the reference, though some shrinking of the uncertainties and mild offsets remain for certain parameter pairs. At $q=16$ (right panel), where the number of simulations greatly exceeds the data-vector dimension, all three estimators produce posteriors that are nearly indistinguishable from the reference, confirming that in the data-rich regime the choice of covariance estimator has a negligible impact on the inferred constraints.

\subsection{Power spectrum results}
\label{sec:Pk_analysis}
\begin{figure}[htbp] 
    \centering
    \includegraphics[width=1\columnwidth]{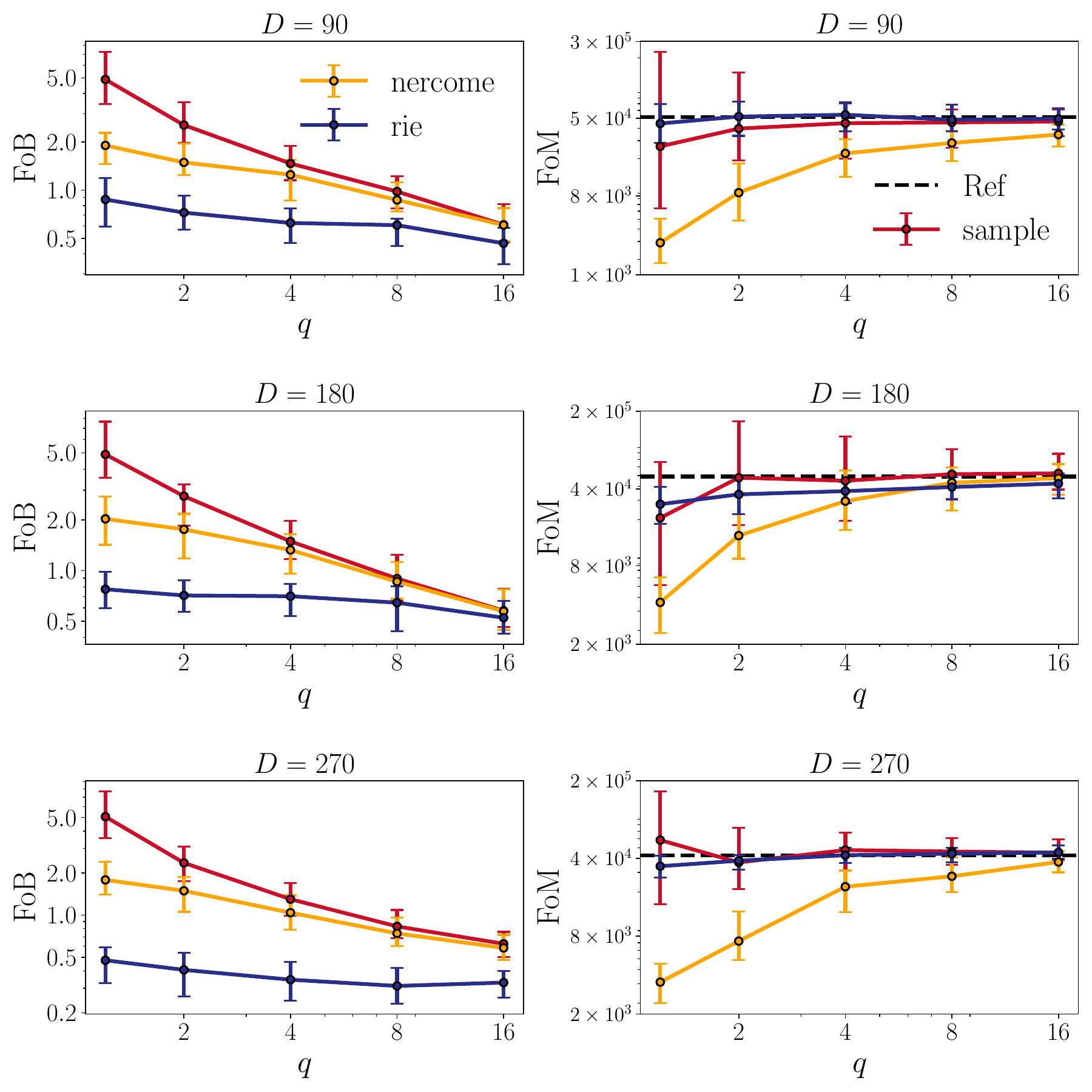}
    \caption{\small \emph{Left panel:} FoB of the power spectrum of the synthetic dataset as a function of the number of mocks used to compute the covariance matrix for the sample (red), NERCOME (gold) and RI (blue) estimators. Dots and errorbars are, respectively, mean and \emph{rms} scatter of the FoB as obtained from 100 different realizations of the covariance matrix. Each row correspond to a different datavector length $D$ as reported in the panel titles. \emph{Right panel:} Same plot but for the FoM. The horizontal, black dashed line represent the FoM associated to the true covariance matrix.}
    \label{fig:FoM_FoB_fourier}
\end{figure}

\begin{figure}[htbp] 
    \centering
    \includegraphics[width=1\columnwidth]{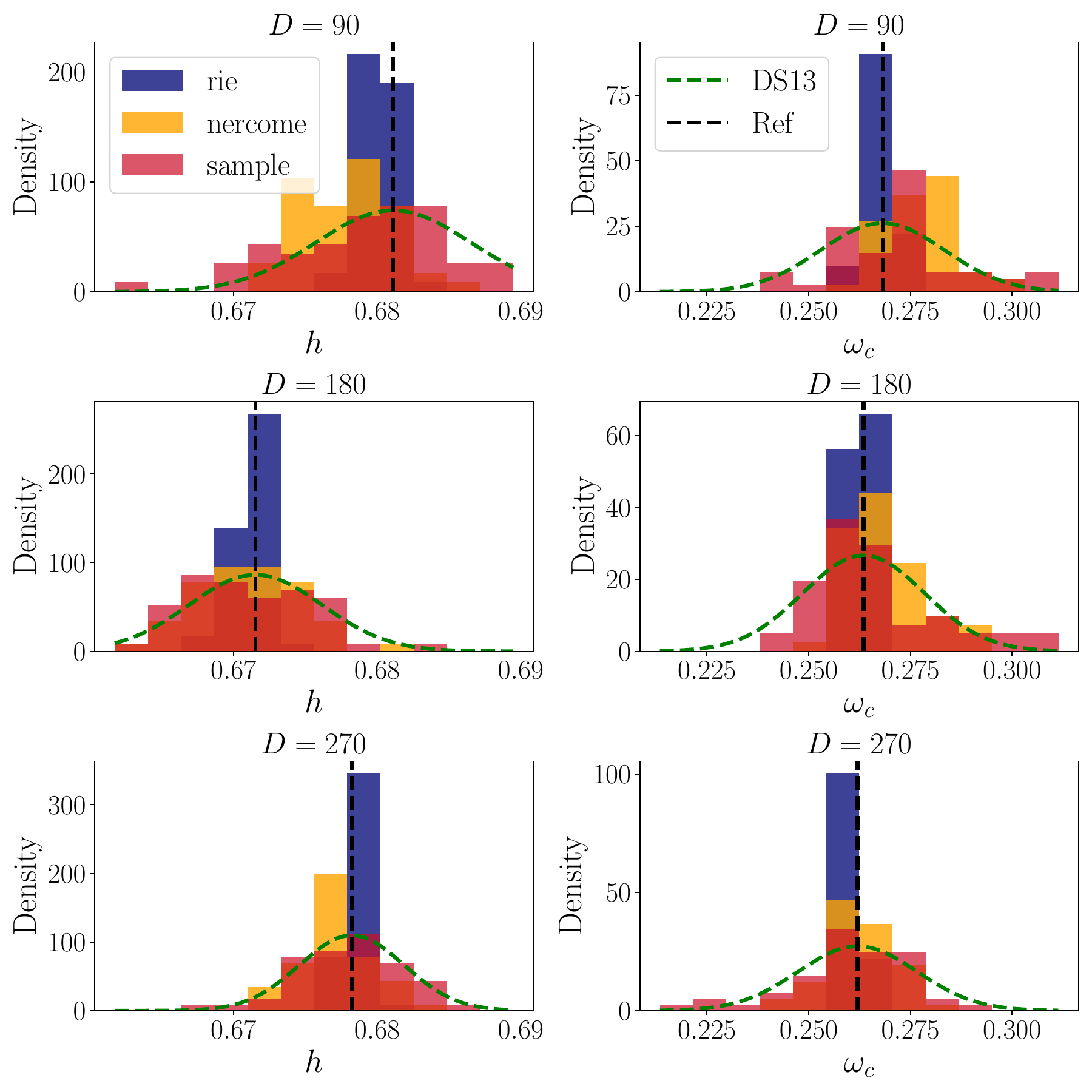}
    \caption{\small Normalised distributions of the best-fit parameter estimates for $h$ (left column) and $\omega_c$ (right column) derived from the power spectrum for different data vector sizes $D$.
    Results are shown for NERCOME (gold), RIE (blue), and sample covariance (red). 
    The vertical black dashed lines indicate the true parameter values inferred from the exact covariance, while the cyan dashed lines show the distribution of best-fit values predicted by DS13.}
    \label{fig:distros_fourier}
\end{figure}

The main results of the power spectrum analysis are summarized in Fig.~\ref{fig:FoM_FoB_fourier} and Fig.~\ref{fig:distros_fourier}, where we maintain the same color scheme as in previous figures. 
In the left panel of Fig.~\ref{fig:FoM_FoB_fourier}, for each of the three $D$ values considered in this work, we present the FoB of the sample, NERCOME and RIE estimators as a function of $q$, along with their respective 68\% uncertainty regions. 

\noindent The behavior of the sample and NERCOME estimators broadly follows the trends observed in configuration space. However, in Fourier space the FoB of the sample covariance starts to deviate significantly from that of NERCOME at an earlier point, already for $q < 4$.
By contrast, the behavior of the RIE estimator is markedly different. Unlike in configuration space, its FoB remains significantly smaller than that of the sample and NERCOME estimators over the entire range of $q$ values explored in this analysis.
Importantly, results are consistent for all the three $D$ values explored, suggesting insensitivity of the method w.r.t. the absolute size of the datavector.\\ 
Additional evidence of the improved accuracy of the RIE estimator in providing reliable best-fit estimates in Fourier space is illustrated in Fig.~\ref{fig:distros_fourier}, which presents the normalized distributions of the parameters $h$ and $\omega_c$ derived setting $q = 2$. Also in this case, the behavior of the remaining free parameters is completely analogous. 

\begin{figure*}[htbp]
    \centering
    \includegraphics[width=\textwidth]{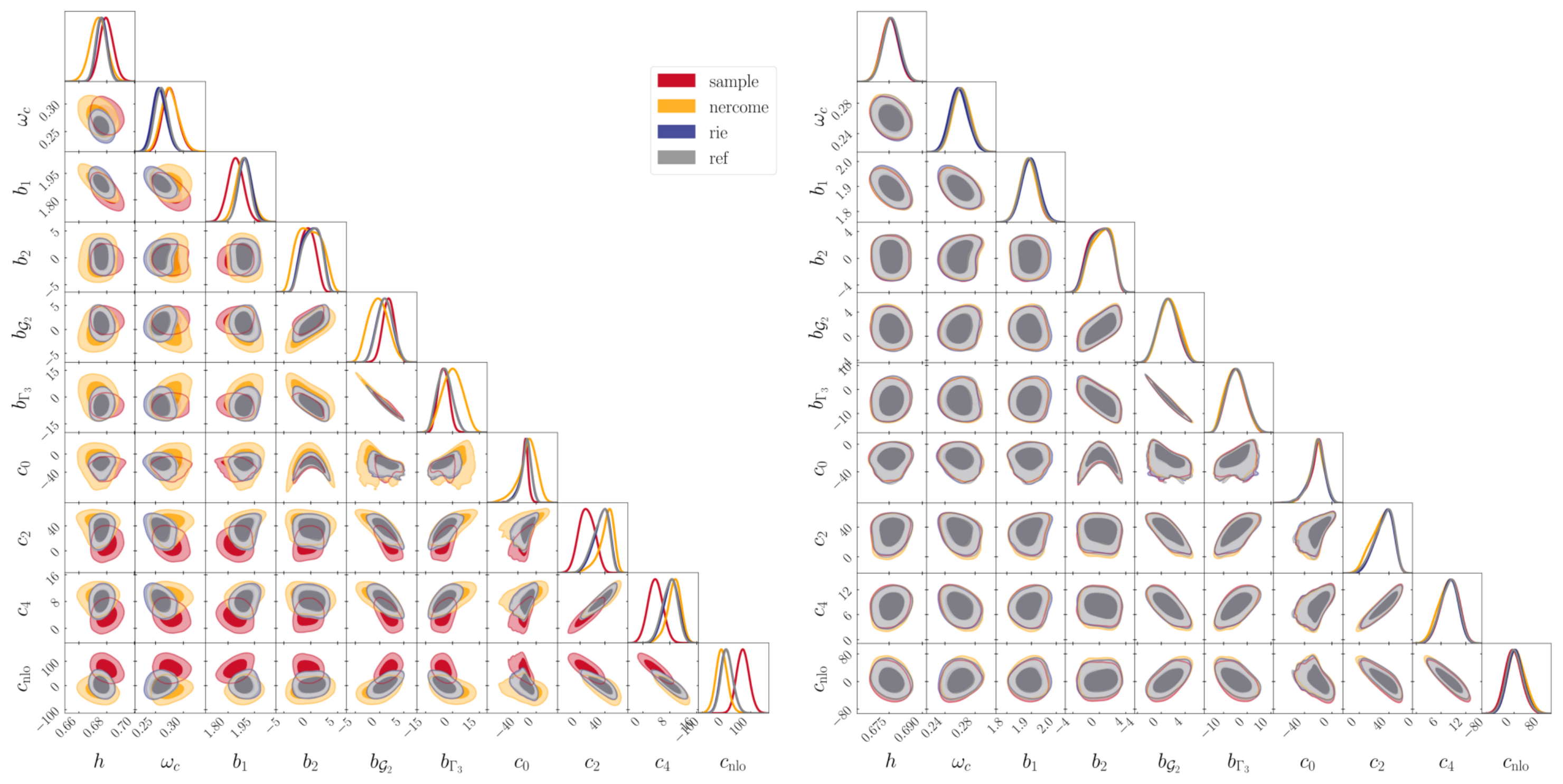}
    \caption{\small Posterior distributions for the model parameters obtained from the power spectrum analysis for $D=270$ using different covariance matrix estimators. For each method, we show the realization whose FoM and FoB are closest to the median values of their respective distributions. Results are shown for two values of $q$ ($q=1.2$ and $q=16$), comparing sample (red), NERCOME (gold), RIE (blue), and the reference case (grey).}
    \label{fig:triangle_fourier_median_real}
\end{figure*}

\noindent Consistent with configuration space results, the best-fit distribution obtained using the sample covariance is broadened according to the DS13 predictions. In contrast, NERCOME and especially RIE provide narrower distributions, confirming that even in Fourier space these estimators are less affected by the Dodelson-Schneider effect (as already shown by \citealt{Beauchamps:2023}). 
In the right panel of Fig.~\ref{fig:FoM_FoB_fourier} we show, as a function of $q$ and for the same datavector sizes, the mean value of the power-spectrum's FoM for the different estimators, along with their 68\% uncertainty regions. Consistent with configuration space, we find that while the sample covariance, when corrected with the Hartlap factor, provides an excellent estimate for the credible volume in parameter space, this is not the case for NERCOME, which exhibits FoMs systematically underestimated w.r.t. the truth (horizontal, black dashed line). This behavior is consistent with the findings of \cite{Beauchamps:2023} and \cite{Looijmans:2024} and, if compared with the configuration space results described in Figs.~\ref{fig:fom_fob_vs_q_multi_bins} and \ref{fig:distros}, suggests that the performance of this estimator is strongly influenced by the structure of the underlying covariance matrix, which is notoriously different in configuration and Fourier space.\\
Similarly, the behavior of the RIE estimator differs markedly from what we observed in configuration space. There, RIE tended to overestimate the FoM, leading to an overly optimistic assessment of parameter constraints. In Fourier space, instead, RIE shows the opposite trend: it yields a mild underestimation of the FoM, i.e. slightly broader posteriors than the reference case, but without the large bias seen in configuration space. \\
This change of behavior further supports the idea that the performance of shrinkage-based estimators such as RIE or NERCOME is highly sensitive to the structure of the true covariance matrix. In particular, the smoother, more diagonally dominated covariance structure typical of power-spectrum data vectors appears to be better matched to the shrinkage model underlying RIE, leading to a much more controlled bias in the inferred FoM.\\
Finally, as already done for the configuration space analysis, we show in Fig.~\ref{fig:triangle_fourier_median_real} the posterior contour plots for the power spectrum case, again selecting for each estimator the realization closest to the median FoM and FoB reported in Fig.~\ref{fig:FoM_FoB_fourier}. Results are shown for $D=270$ with $q=1.2$ (left) and $q=16$ (right).

\noindent The picture that emerges is markedly different from the configuration-space case. For $q=1.2$, the sample (red) and NERCOME (gold) contours are both significantly broader than the reference (grey), with NERCOME now exhibiting a level of degradation comparable to the sample estimator. This is in stark contrast to the 2PCF results, where NERCOME performed considerably better. Several NERCOME marginals appear visibly inflated, consistent with the overestimation of uncertainties reported in previous power spectrum studies. The match between the RIE contours (blue) and the reference ones is instead striking: even at $q=1.2$, the the RIE closely trace the reference posteriors across all parameter pairs, both in shape and position, demonstrating a remarkable robustness to the low-simulation regime. For $q=16$ (right panel), all estimators converge toward the reference, as expected. These results reinforce and visually complement the quantitative findings of this section, highlighting how the relative performance of covariance estimators can depend significantly on the summary statistic under consideration.
 
\section{Conclusions}
\label{sec:conclusion}

In this work we investigated whether denoising techniques rooted in Random Matrix Theory can stabilize cosmological parameter inference from galaxy clustering when the number of mock realizations available for covariance estimation is comparable to the data-vector dimension. To this end, we built a controlled framework based on synthetic 2PCF and power-spectrum measurements with analytically known covariance matrices, and compared three estimators: the standard sample covariance (with the Hartlap finite-$N$ correction), the non-linear shrinkage estimator NERCOME, and the Rotational Invariant Estimator (RIE) applied to the correlation matrix. The latter estimator is employed here for the first time in the context of cosmology.\\
Our analysis reveals a clear hierarchy among the three methods. The sample covariance, once corrected for the known bias in the precision matrix, recovers accurate average posterior volumes across all regimes explored. However, as the ratio $q = N/D$ decreases, sampling noise induces increasingly large stochastic shifts in the best-fit parameters, consistent with the Dodelson--Schneider effect, making the sample estimator unreliable for parameter estimation in the low-mock regime.
Both NERCOME and RIE substantially suppress these stochastic shifts, yielding narrower best-fit distributions and smaller FoB values than the sample estimator. However, their impact on the inferred posterior volume is probe-dependent. In configuration space, both estimators tend to produce overly tight constraints, with a FoM bias that grows with the data-vector dimension $D$. In Fourier space, the picture changes significantly: NERCOME overestimates the constraining power, yielding inflated posterior volumes in agreement with the findings of \cite{Beauchamps:2023} and \cite{Looijmans:2024}, while RIE stands out as the best-performing estimator, achieving the strongest suppression of best-fit scatter across all $q$ and $D$ values explored and recovering posterior volumes consistent with the reference. Even at $q = 1.2$, its posteriors closely trace the reference contours in both shape and position, a level of robustness no other estimator in our comparison matches. In this regard, we note that the covariance structure of Fourier-space observables is typically smoother and more diagonally dominated than its configuration-space counterpart, a property that likely plays a role in the superior performance of RIE, though a rigorous understanding of this connection remains to be established.\\
These findings open several avenues for future work. First, validating RIE on more realistic covariance models, including non-Gaussian contributions, survey geometry effects, and observational systematics, will be essential before deploying it in the analysis of real data. Second, the strong stabilization of best-fit parameters suggests that hybrid strategies, combining RIE-based denoising to suppress best-fit wander with targeted calibration of the posterior volume, may offer a practical path for future large-scale structure analyzes.
Perhaps most importantly, the covariance estimation challenge is expected to become dramatically more severe for higher-order clustering statistics. For the bispectrum or the three-point correlation function, the data-vector grows rapidly with the number of triangle configurations, easily reaching $\mathcal{O}(10^2)$ bins \citep{ Slepian:2017B, Slepian:2017L, Pearson:2018, Sugiyama:2019, Guidi:2022, Rizzo:2023, Labate:2025}, and the number of mocks required for a stable sample covariance estimate quickly becomes prohibitive \citep[see e.g.][]{Colavincenzo:2020, Veropalumbo:2022}. The same applies to joint analyses that combine two-point and higher-order statistics, where the concatenated data vector further exacerbates the dimensionality problem \citep{GilMarin:2014, GilMarin:2016, Sugiyama:2019, Oddo:2020, Oddo:2021, Sugiyama:2021, Veropalumbo:2021,  D'Amico:2022a, Farina:2026, Guidi:2026, Pardede:2026}. The ability of RIE to provide reliable parameter inference even when $q$ is close to unity makes it a natural candidate for these applications, where the limitations of conventional sample covariance estimation are most acute. Extending the present analysis to these regimes represents a compelling next step.

\begin{acknowledgements}
The authors thank E. Branchini, J. Bun, M. Raveri and M. Moresco for useful discussions and comments on the results.
Simulations and computations in this work have been
run at the computing facilities of INFN, Sezione di Genova: the authors wish to
thank the INFN IT personnel in Genova for their precious and constant support.
This work was supported by the ASI/INAF agreement n. 2018-23-HH.0 “Scientific activity for Euclid mission, Phase D”, the research grant ‘From Darklight to DM: understanding the galaxy/matter connection to measure the Universe’, the INFN project “InDark”. MG acknowledges the financial contribution from the grant PRIN-MUR 2022 2022NY2ZRS 001 “Optimizing the extraction of cosmological information from Large Scale Structure analysis in view of the next large spectroscopic surveys” supported by Next Generation EU.
\end{acknowledgements}

%
%

\bibliographystyle{aa}
\bibliography{a3PCF.bib}


\begin{appendix}
\section{2PCF and power spectrum emulators}
\label{app:emulator}

To emulate the galaxy power spectrum and 2PCF multipoles, we employ a fully connected feed-forward neural network for each of the 18 Effective Field Theory (EFT) basis functions  \citep[see][]{Farina:2026} as a function of
the cosmological parameters $h$ and $\omega_c$. The predicted basis functions
are then contracted with the EFT bias and counterterm coefficients at each step
of the MCMC chain, which allows a single set of trained networks to serve any
combination of nuisance parameters without retraining.

\paragraph{Input preprocessing.}
The two raw cosmological inputs $(h, \omega_c)$ are first min-max normalised to
$[0,1]$ using the training-set extrema and then mapped to a five-dimensional
feature vector via a second-order polynomial expansion,

\begin{equation}
  \boldsymbol{\phi}(h,\omega_c)
  =
  \bigl[\tilde{h},\,\tilde{\omega}_c,\,\tilde{h}^{2},\,
        \tilde{\omega}_c^{2},\,\tilde{h}\,\tilde{\omega}_c\bigr]^{\!\top},
  \label{eq:feature_expansion}
\end{equation}

\noindent
where $\tilde{h}=(h-h_{\min})/(h_{\max}-h_{\min})$ and analogously for
$\tilde{\omega}_c$. This quadratic expansion enriches the input representation
and improves accuracy relative to passing the raw parameters directly into the
network.

\paragraph{Architecture.}
The network consists of six fully connected layers. The first hidden layer
projects the five input features to 256 neurons. Four subsequent hidden layers
maintain a constant width of 256 neurons each and include residual (skip)
connections,

\begin{equation}
  \mathbf{z}^{(l+1)} = f\!\left(W^{(l)}\mathbf{z}^{(l)} + \mathbf{b}^{(l)}\right)
                        + \mathbf{z}^{(l)},
  \qquad l = 2,\ldots,5,
  \label{eq:residual}
\end{equation}

\noindent
where $W^{(l)}$ and $\mathbf{b}^{(l)}$ are the weight matrix and bias vector of
layer $l$. The output layer maps the 256-dimensional representation to the
flattened target of dimension $3\times18\times N_{\rm bins}$, where the three
indices correspond to the monopole ($\ell=0$), quadrupole ($\ell=2$), and
hexadecapole ($\ell=4$), respectively. The first hidden layer uses no skip
connection because the input and hidden dimensions differ.
Each hidden layer is followed by the parametrised activation function introduced
in \citet{Alsing:2020,DeRose:2022},

\begin{equation}
  f(x;\beta,\gamma)
  =
  \Biggl(\gamma + \frac{1-\gamma}{1 + e^{-\beta x}}\Biggr)\cdot x,
  \label{eq:derose}
\end{equation}

\noindent
where $\beta$ and $\gamma$ are scalar learnable parameters, independent for
each layer, that control the steepness and baseline of the sigmoid gate. This
activation is more expressive than a fixed nonlinearity and can smoothly
interpolate between linear ($\gamma\to 1$) and gated-linear behaviour.

\paragraph{Training.}
The network is trained on $10\,000$ samples drawn via Latin Hypercube Sampling
(LHS) and validated on a held-out set of $2\,500$ samples covering the same
cosmological parameter space. We minimize the mean squared error (MSE) on the
standardised outputs using the Adam optimiser \citep{Kingma:2014} with
$\beta_1=0.9$, $\beta_2=0.999$, and $\varepsilon=10^{-8}$.\\
Training proceeds in four sequential phases, each with its own peak learning
rate $\eta_{\max}$ and mini-batch size, following a cosine-annealing schedule
within each phase,

\begin{equation}
  \eta(e)
  =
  \eta_{\min}
  + \frac{1}{2}\!\left(\eta_{\max} - \eta_{\min}\right)
    \!\left(1 + \cos\!\frac{\pi\, e}{E}\right),
  \label{eq:lr_schedule}
\end{equation}

\noindent
where $e$ is the epoch index within the phase, $E$ is the total number of
epochs in that phase, and $\eta_{\min}=\eta_{\max}/100$. The four phases are
summarised in Table~\ref{tab:training_schedule}.

\begin{table}[h]
  \centering
  \caption{Multi-stage training schedule used for the emulators adopted in this work.}
  \label{tab:training_schedule}
  \begin{tabular}{ccccc}
    \hline\hline
    Phase & $\eta_{\max}$ & Epochs & Batch size & Purpose \\
    \hline
    1 & $5\times10^{-3}$ & 80 & 256 & Coarse exploration \\
    2 & $1\times10^{-3}$ & 80 & 128 & Medium refinement \\
    3 & $3\times10^{-4}$ & 80 &  64 & Fine tuning \\
    4 & $5\times10^{-5}$ & 60 &  32 & Convergence \\
    \hline
    \multicolumn{2}{l}{Total} & 300 & & \\
    \hline\hline
  \end{tabular}
\end{table}

\noindent We assessed the accuracy of the 18 trained emulators on the 2,500-sample test set by evaluating the fully reconstructed observables, obtained by resumming the emulator outputs using the fiducial nuisance parameters listed in Tab.~\ref{tab:nuisance_pars}. In this way, the validation is performed directly on the final predictions for the power spectrum multipoles and 2PCF, rather than on the individual emulator components.
For each test cosmology and each scale bin, we computed the percentage difference between the emulated and the true observable. For both the power spectrum and the 2PCF multipoles, the median emulation error is at the percent level across all multipoles and scales considered, well within the statistical uncertainties explored in this paper.

\end{appendix}

\label{LastPage}
\end{document}